\documentstyle[fleqn,epsf]{article}
\def\beq{\begin{equation}}
\def\eeq{\end{equation}}
\def\Eq{Eq.~(\ref}
\def\sfA{{\sf A}}
\def\sfB{{\sf B}}
\def\sfC{{\sf C}}
\def\0{\otimes}
\def\6{\langle}
\def\9{\rangle}
\def\Tr{\mbox{Tr}\,}
\def\bc{\medskip\begin{center}}
\def\ec{\end{center}\medskip}

\begin{document}

\renewcommand{\thefootnote}{\fnsymbol{footnote}}

\vspace*{\fill}
\begin{center}
{\Large {\bf Classical interventions in quantum systems.\medskip 

II. Relativistic invariance}}\\[15mm]

Asher Peres\footnote{E-mail: peres@photon.technion.ac.il} \\[8mm]
{\sl Department of Physics, Technion---Israel Institute of Technology,
32\,000 Haifa, Israel}\\[15mm]

{\bf Abstract}\end{center}

If several interventions performed on a quantum system are localized in
mutually space-like regions, they will be recorded as a sequence of
``quantum jumps'' in one Lorentz frame, and as a different sequence of
jumps in another Lorentz frame. Conditions are specified that must be
obeyed by the various operators involved in the calculations so that
these two different sequences lead to the same observable results. These
conditions are similar to the equal-time commutation relations in
quantum field theory. They are sufficient to prevent super\-luminal
signaling. (The derivation of these results does not require most of the
contents of the preceding article. What is needed is briefly summarized
here, so that the present article is essentially self-contained.)\vfill

\noindent PACS numbers: 03.65.Bz, \ 03.30.+p, \ 03.67.* \vfill

\noindent Physical Review A {\bf61}, 022117 (2000)\vfill

\newpage \begin{center}{\bf I. THE PROBLEM}\ec

Quantum measurements~\cite{WZ} are usually considered as
quasi-instantaneous processes. In particular, they affect the wave
function instantaneously throughout the entire configuration space.
Measurements of finite duration~\cite{PW85} cannot alleviate this
conundrum. Is this quasi-instantaneous change of the quantum state,
caused by a local intervention, consistent with relativity theory? The
answer is not obvious. The wave function itself is not a material object
forbidden to travel faster than light, but we may still ask how the
dynamical evolution of an extended quantum system that undergoes several
measurements in distant spacetime regions is described in different
Lorentz frames. 

Difficulties were pointed out long ago by Bloch~\cite{Bloch}, Aharonov
and Albert~\cite{AA}, and many others~\cite{NYAS}. Still before them,
in the very early years of quantum mechanics, Bohr and
Rosenfeld~\cite{BR33} had given a complete relativistic theory of the
measurement of quantum {\it fields\/}, but these authors were not
concerned about the properties of the new quantum states that resulted
from these measurements and their work does not answer the question that
was raised above. Other authors~\cite{qt,SS97} considered detectors in
relative motion, and therefore at rest in different Lorentz frames.
These works also do not give an explicit answer to the above question: a
detector in uniform motion is just as good as one that has undergone an
ordinary spatial rotation (accelerated detectors involve new physical
phenomena~\cite{Unruh} and are not considered in this article). The
point is not how individual detectors happen to move, but how the
effects due to these detectors are described in different ways in one
Lorentz frame or another.

In the preceding article~\cite{I}, the notion of measurement was
extended to the more general one of {\it intervention\/}. An
intervention consists of the acquisition and recording of information by
a measuring apparatus, possibly followed by the emission of classical
signals for controlling the execution of further interventions. More
generally, a consequence of the intervention may be a change of the
environment in which the quantum system evolves. These effects are the
{\it output\/} of the intervention. These notions are refined in
Sect.~II of the present article so as to be applicable to relativistic
situations.

A relativistic treatment is essential to analyze space-like separated
interventions, such as in Bohm's version of the Einstein-Podolsky-Rosen
``paradox'' \cite{EPR,Bohm} (hereafter EPRB) which is sketched in
Fig.~1, with two coordinate systems in relative motion. In that
experiment, a pair of spin-$1\over2$ particles is prepared in a singlet
state at time $t_0$ (referred to one Lorentz frame) or $t'_0$ (referred
to another Lorentz frame). The particles move apart and are detected by
two observers. Each observer measures a spin component along an
arbitrarily chosen direction. The two interventions are mutually
space-like as shown in the figure. Event \sfA\ occurs first in $t$-time,
and event \sfB\ is the first one in $t'$-time. The evolution of the
quantum state of this bipartite system appears to be genuinely different
when recorded in two Lorentz frames in relative motion. The quantum
states are not Lorentz-transforms of each other. Yet, all the observable
results are the same. Consistency of the theoretical formalism imposes
definite relationships between the various operators used in the
calculations. These are investigated in Sect.~III.

Another example, this one taken from real life, is the detection system
in the experimental facility of a modern high energy
accelerator~\cite{LEP}. Following a high energy collision, thousands of
detection events occur in locations that may be mutually space-like.
Yet, some of the detection events are mutually time-like, for example
when the world line of a charged particle is recorded in an array of
wire chambers. High energy physicists use a language which is different
from the one in the present article. For them, an ``event'' is one high
energy collision together with all the subsequent detections that are
recorded. This ``event'' is what I call here an experiment (while they
call ``experiment'' the complete experimental setup that may be run for
many months). Their ``detector'' is a huge machine weighing thousands of
tons, while here the term detector means each elementary detecting
element, such as a new bubble in a bubble chamber or a small segment of
wire in a wire chamber. (A typical wire chamber records only which wire
was excited. However, it is in principle possible to approximately
locate the place in that wire where the electric discharge occurred, if
we wish to do so.) Apart from the above differences in terminology, the
events that follow a high energy collision are an excellent example of
the circumstances discussed in the present article.

Returning to the Einstein-Podolsky-Rosen conundrum, we must analyze
whether it actually involves a genuine quantum nonlocality. Such a claim
has led some authors to suggest the possibility of superluminal
communication. This would have disastrous consequences for relativistic
causality~\cite{jossi}. Bell's theorem~\cite{Bell} asserts that it is
impossible to mimic quantum correlations by classical local ``hidden''
variables, so that any classical imitation of quantum mechanics is
necessarily nonlocal. However Bell's theorem does not imply the
existence of any nonlocality in quantum theory itself. It is shown in
Sect.~IV that quantum measurements do not allow any information to be
transmitted faster than the characteristic velocity that appears in the
Green's functions of the particles involved in the experiment. In a
Lorentz invariant theory, this limit is the velocity of light, of
course. The last section is devoted to a few concluding remarks.

\bc {\bf II. RELATIVISTIC INTERVENTIONS}\ec

This section includes a brief summary of some parts of the preceding
article~\cite{I} and contains all the material necessary to make the
present one self-contained. Besides this summary, new notions are
introduced to cope with the relativistic nature of the phenomena under
discussion.

First, recall that each intervention is described by a set of {\it
classical\/} parameters~\cite{BJ,percival}. The latter include the
location of that intervention in spacetime, referred to an arbitrary
coordinate system. The coordinates are classical numbers, just as time
in the Schr\"odinger equation is a classical parameter. We also have to
specify the speed and orientation of the apparatus in that coordinate
system and various other {\it input\/} parameters that define the
experimental conditions under which the measuring apparatus operates.
The input parameters are determined by classical information received
from the past light-cone at the point of intervention, or they may be
chosen arbitrarily (in a random way) by the observer and/or the
apparatus.

I just mentioned the existence of a past light-cone. Actually, the only
notion needed at the present stage is a {\it partial ordering\/} of the
interventions: namely, there are no closed causal loops. This property
defines the terms earlier and later. The input parameters of an
intervention are deterministic (or possibly stochastic) functions of the
parameters of earlier interventions, but not of the stochastic outcomes
resulting from later interventions, as explained below.

In the conventional presentation of non-relativistic quantum mechanics,
each intervention has a (finite) number of {\it outcomes\/}, which are
also known as ``results of measurements'' (for example, this or that
detector clicks). In a relativistic treatment, the spatial separation of
the detectors is essential and each detector corresponds to a different
intervention. The reason is that if several detectors are set up so that
they act at a given time in one Lorentz frame, they would act at
different times in another Lorentz frame. However, a knowledge of the
time ordering of events is essential in our dynamical calculations, so
that we want the parameters of an intervention to refer unambiguously to
only one time (indeed to only one spacetime point). Therefore, an
intervention can involve only one detector and it can have only two
possible outcomes: either there was a ``click'' or there wasn't.

Note that the {\it absence\/} of a click, while a detector was present,
is also a valid result of an intervention. The state of the quantum
system does not remain unchanged: it has to change to respect unitarity.
The mere presence of a detector that could have been excited implies
that there has been an interaction between that detector and the quantum
system. Even if the detector has a finite probability of remaining in
its initial state, the quantum system correlated to the latter acquires
a different state~\cite{Dicke}. The absence of a click, when there could
have been one, is also an event and is part of the historical record.

The effect of an intervention on a quantum system initially prepared in
the state $\rho$ is given by Eq.~(20) in the preceding article:

\beq \rho\to\rho'_\mu=\sum_m A_{\mu m}\,\rho\,A_{\mu m}^\dagger,
 \label{ArhoA}\eeq
where $\mu$ is a label that indicates which detector was involved and
whether or not it was activated. The initial $\rho$ is assumed to be
normalized to unit trace, and the trace of $\rho'_\mu$ is the
probability of occurrence of outcome $\mu$. Each symbol $A_{\mu m}$  in
the above equation represents a {\it matrix\/} (not a matrix element).
These may be rectangular matrices where the number of rows depends on
$\mu$. The number of columns is of course equal to the order of the
initial $\rho$. Thus, the Hilbert space of the resulting quantum system
may have a different number of dimensions than the initial one. A
quantum system whose description starts in a given Hilbert space may
evolve in a way that requires a set of Hilbert spaces with different
dimensions. If one insists on keeping the same Hilbert space for the
description of the entire experiment, with all its possible outcomes,
this can still be achieved by defining it as a Fock space.

Each experiment yields a {\it record\/} that comprises a complete list
of which detectors were available (including when and where) and whether
these detectors reacted. Such a record is objective: everyone agrees on
what happened (e.g., which detectors clicked) irrespective of the state
of motion of the observers who read these records. Therefore everyone
agrees on the relative frequency of each type of record among all the
records that are observed if the experiment is repeated many times, and
the theoretical probabilities also have to be the same for everyone.

What is the role of relativity theory here? We may likewise ask what is
the role of translation and/or rotation invariance in a nonrelativistic
theory. The point is that the rules for computing quantum probabilities
involve explicitly the spacetime coordinates of the interventions.
Lorentz invariance (or rotation invariance, as a special case) says that
if the classical spacetime coordinates are subjected to a particular
linear transformation, then the probabilities remain the same. This
invariance is not trivial because the rule for computing the probability
of occurrence of a given record involves a sequence of mathematical
operations corresponding to the time ordered set of all the relevant
interventions. If we only consider the Euclidean group, all we have to
know is how to transform the classical parameters, and the wave
function, and the various operators, under translations and rotations of
the coordinates. However, when we consider genuine Lorentz
transformations, we have not only to Lorentz-transform the above
symbols, but we are faced with a new problem: the natural way of
calculating  the result of a sequence of interventions, namely by
considering them in chronological order, is different for different
inertial frames. The issue is not only a matter of covariance of the
symbols at each intervention and between consecutive interventions.
There are genuinely different prescriptions for choosing the sequence of
mathematical operations in our calculation. The principle of relativity
asserts that there are no privileged inertial frames. Therefore these
different orderings ought to give the same set of probabilities, and
this demand is not trivial.

The experimental records are the only real thing we have to consider.
Their observed relative frequencies are objective numbers and are
Lorentz invariant. On the other hand, wave functions and operators are
mathematical concepts useful for computing quantum probabilities, but
they have no real existence~\cite{Stapp}. All the difficulties that have
been associated with a relativistic theory of quantum measurements are
due to attributing a real nature to the symbols that represent quantum
states.

Note also that while interventions are localized in spacetime, quantum
systems are pervasive. In each experiment, irrespective of its history,
there is only one quantum system. The latter typically consists of
several particles or other subsystems, some of which may be created or
annihilated by the various interventions. The next two sections of this
article are concerned with sharp localized interventions on quantum
systems that freely evolve throughout spacetime between these
interventions, and in particular with the Lorentz covariance of the
results.

\bc {\bf III. TWO MUTUALLY SPACELIKE INTERVENTIONS}\ec

Consider again the EPRB gedankenexperiment which is depicted in Fig.~1,
with two coordinate systems in relative motion. There exists a Lorentz
transformation connecting the initial states $\rho$ (at time $t_0$) and
$\rho'$ (at time $t_0'$) before the two interventions, and likewise
there is a Lorentz transformation connecting the final states at times
$t_f$ and $t_f'$ after completion of the two interventions. On the other
hand, there is no Lorentz transformation relating the states at
intermediate times represented by the lines that pass between
interventions \sfA\ and \sfB~\cite{Bloch,AA}. This may be contrasted
with the ontology of classical relativistic theory. Classical theory
asserts that fields, velocities, etc., transform in a definite way and
that the equations of motion of particles and fields behave covariantly.
For example if the expression for the Lorentz force is written
$f_\mu=F_{\mu\nu}u^\nu$ in one frame, the same expression is valid in
any other frame. These symbols ($f_\mu$, etc.) have objective values.
They represent entities that really exist, according to the theory. On
the other hand, wave functions have no objective value. They do not
transform covariantly when there are interventions. Only the classical
parameters attached to each intervention transform covariantly. Yet, in
spite of the non-covariance of $\rho$, the final results of the
calculations (the probabilities of specified sets of events) are Lorentz
invariant.

Note that each line in Fig.~1 represents one instant of the time
coordinate, as in the ordinary non-relativistic formulation of quantum
mechanics. There is no way of defining a relativistic proper time for a
quantum system which is spread all over space. It is possible to define
a proper time for each {\it apparatus\/}, which has classical
coordinates and follows a continuous world-line. However, this is not
necessary. We are only interested in a {\it discrete\/} set of
interventions, and the latter are referred to a common coordinate system
that covers the whole of spacetime. There is no role for the private
proper times that might be attached to the apparatuses' world-lines.

If we attempt to generalize the parallel straight lines in Fig.~1 to a
spacelike foliation in a curved spacetime, as we would have in general
relativity, we encounter the difficulty that no such foliation may exist
globally. However, there is no need for such a global foliation and in
particular we do not assume the validity of a Schwinger-Tomonaga
equation, $i\delta{\mit\Psi}/\delta\sigma=H(\sigma)
{\mit\Psi}$, as can be found in the work of Aharonov and
Albert~\cite{AA}. The only condition that we need is the absence of
closed timelike curves. Namely, if two events can be connected by
continuous timelike (or null) curves, without past-future zigzags, then
all these curves have the same orientation.

Returning to special relativity, consider the evolution of the quantum
state in the Lorentz frame where intervention \sfA\ is the first one to
occur and has outcome $\mu$, and \sfB\ is the second intervention, with
outcome $\nu$. Between these two events, nothing actually happens in the
real world. It is only in our mathematical calculations that there is a
deterministic evolution of the state of the quantum system. This
evolution is {\it not\/} a physical process. For example, the quantum
state of Schr\"odinger's legendary cat, doomed to be killed by an
automatic device triggered by the decay of a radio\-active atom, evolves
into a superposition of ``live'' and ``dead'' states. This is a
manifestly absurd situation for a real cat. The only meaning that such a
quantum state can have is that of a mathematical tool for statistical
predictions on the fates of numerous cats subjected to the same cruel
experiment. 

What distinguishes the intermediate evolution {\it between\/}
interventions from the one occurring {\it at\/} an intervention is the
unpredictability of the outcome of the latter: either there is a click
or there is no click of the detector. This unpredictable macroscopic
event starts a new chapter in the history of the quantum system which
acquires a new state, according to \Eq{ArhoA}). As long as there is no
such branching, the quantum evolution will be called {\it free\/}, even
though it may depend on external classical fields that are specified by
the classical parameters of the preceding interventions.

Quantum mechanics asserts that during the free evolution of a closed
quantum system, its state undergoes a unitary transformation generated
by a Hamiltonian. The latter depends in a prescribed way on the
preceding outcome(s) according to the protocol that has been specified
for the experiment. The unitary operator for the evolution following
intervention \sfA\ with outcome $\mu$, and ending at intervention \sfB,
will be denoted by $U_{BA_\mu}$. (More generally, it is possible to
consider an evolution which is continuously perturbed by the
environment, as in the last section of the preceding article~\cite{I}.
In that case, the unitary evolution would be replaced by a more general
continuous completely positive map, so that instead of $U_{BA_\mu}$
there would be Kraus operators with additional indices to be summed
over. I shall refrain from using this more general formalism so as not
to get into an unnecessarily complicated argument. Anyway, the presence
of such a pervasive environment would break Lorentz invariance.)

Note that the chronological order of the indices in $U_{BA_\mu}$ is from
right to left (just as is the order for consecutive applications of a
product of linear operators), and in particular that $U_{BA_\mu}$ does
not depend on the future outcome at intervention \sfB. Likewise, there
is a unitary operator $U_{A0}$ for the evolution that precedes event
\sfA, and an operator $U_{fB_\nu}$ for the final evolution that follows
outcome $\nu$ of intervention \sfB. The final quantum state at time
$t_f$ is given by a generalization of \Eq{ArhoA}):

\beq \rho_f=\sum_{m,n} K_{mn}\,\rho\,K_{mn}^\dagger, \label{rhof}\eeq
where

\beq K_{mn}=U_{fB_\nu}\,B_{\nu n}\,U_{BA_\mu}\,A_{\mu m}\,U_{A0}.
 \label{K} \eeq

The same events can also be described in the Lorentz frame where \sfB\
occurs first. We have, with the primed variables,

\beq \rho'_f=\sum_{m,n} L'_{mn}\,\rho'\,L_{mn}'^\dagger, \eeq
where

\beq L'_{mn}=
  V'_{fA_\mu}\,A'_{\mu m}\,V'_{AB_\nu}\,B'_{\nu n}\,V'_{B0}. \eeq
Here, the unitary operator for the free evolution between the two
interventions has been denoted by $V'_{AB_\nu}$. It is not related in
any obvious way to the operator $U_{BA_\mu}$. These operators indeed
correspond to different slabs of spacetime. Likewise the other evolution
operators in the primed coordinates have been called $V'$ with
appropriate subscripts. Note that $\Tr(\rho_f)=\Tr(\rho'_f)$ is the
joint probability of occurrence of the records $\mu$ and $\nu$ during
the experiment.

Einstein's principle of relativity asserts that there is no privileged
inertial frame, and therefore both descriptions given above are equally
valid. Formally, the states $\rho_f$ (at time $t_f$) and the state
$\rho'_f$ (at time $t'_f$) have to be Lorentz transforms of each other.
This requirement imposes severe restrictions on the various matrices
that appear in the preceding equations. In order to investigate this
problem, consider a continuous Lorentz transformation from the primed
to the unprimed frame. As long as the order of occurrence of \sfA\ and
\sfB\ is not affected by this continuous transformation of the spacetime
coordinates, the latter is implemented in the quantum formalism by
unitary transformations of the various operators. These unitary
transformations obviously do not affect the observable probabilities.

Therefore, in order to investigate the issue of relativistic invariance,
it is sufficient to consider two Lorentz frames where \sfA\ and \sfB\
are almost simultaneous: either \sfA\ occurs just before \sfB, or just
after \sfB. There is of course no real difference in the actual physical
situations and the Lorentz ``transformation'' between these two
arbitrarily close frames (primed and unprimed) is performed by the unit
operator. In particular, $U_{BA_\mu}={\bf1}=V'_{AB_\nu}$, since there is
no finite time lapse for any evolution to occur between the two events.
The only difference resides in our method for calculating the final
quantum state: first \sfA\ then \sfB, or first \sfB\ then \sfA.
Consistency of the two results is obviously achieved if

\beq A_{\mu m}\,B_{\nu n}=B_{\nu n}\,A_{\mu m}, \label {ABBA}\eeq
or 
\beq [A_{\mu m}, B_{\nu n}]=0. \label{etcr} \eeq
This equal-time commutation relation, which was derived here as a
sufficient condition for consistency of the calculations, is always
satisfied if the operators $A_{\mu m}$ and $B_{\nu n}$ are direct
products of operators pertaining to the two subsystems:

\beq A_{\mu m}=a_{\mu m}\0{\bf1} \qquad\mbox{and}\qquad
 B_{\nu n}={\bf1}\0b_{\nu n}, \label{dirprod} \eeq
where {\bf1} now denotes the unit matrix of each subsystem. This
relationship is obviously fulfilled if there are two distinct
apparatuses whose dynamical variables commute, and moreover if {\it the
dynamical variables of the quantum subsystems commute\/}. This is indeed
a necessary condition for legitimately calling them subsystems.

The analogy with relativistic quantum field theory is manifest: field
operators belonging to points at space-like distances commute (or
anti\-commute in the case of fermionic fields). Quantum field theory
mostly uses the Heisenberg picture or the interaction picture, while in
the present work it is the Schr\"odinger picture that is employed. This
makes no difference in \Eq{etcr}) which applies to equal times. Could we
have here too anti\-commutation relations? It is easily seen that it is
possible to introduce a minus sign on the right hand side of \Eq{ABBA}),
or even an arbitrary phase factor $e^{i\phi_{AB}}$. However, this
generalization will not be investigated in the present article whose
subject is quantum mechanics, not quantum field theory.
 
One may wonder whether the result expressed in \Eq{dirprod}) is trivial.
Direct products were postulated in the very early years of quantum
mechanics by Weyl~\cite{weyl} as the only reasonable way for describing
composite systems. Here, this representation was derived from an
argument involving Lorentz invariance. However, such a proof may well be
circular~\cite{kennedy}: it assumes a relativistic partial ordering of
events, i.e., the impossibility of super\-luminal signaling, while this
impossibility is proved in quantum field theory by assuming the tensor
product representation for composite systems. This issue was also
investigated by Rosen~\cite{rosen} in the context of molecular biology.
According to Rosen, while any micro\-physical system can be expressed as
a composite of subsystems, there is no reason to suppose that such a
factorization is unique, because rings of operators may in general be
factored in many distinct ways. Only if it were found that the
factorization is unique, this would imply that there is only one way in
which the state of a system can be synthetized from the states of
simpler subsystems.

Returning to \Eq{dirprod}), it is important to remember that an
intervention can change the dimensions of the quantum system. Here is a
simple example. The quantum system initially consists of a pair of
spin-$1\over2$ particles, as in the EPRB experiment. The two observers
are called Alice and Bob, as usual. Alice, who intervenes at \sfA, uses
an apparatus that contains a subsystem $\cal S$ prepared as an entangled
state of a spin-$1\over2$ particle and a particle of spin~1. She
receives a particle of spin~$1\over2$ (that is, one of the two particles
of the quantum system under observation) and she measures the Bell
operator~\cite{BMR} of the composite system formed by that particle and
the spin-$1\over2$ particle in~$\cal S$. That measurement can have four
different outcomes, and according to its result Alice performs one of
four specified unitary rotations on the spin~1 particle of~$\cal S$. She
then discards everything but that particle of spin~1, and she releases
the latter for future experiments. In this way, Alice's intervention
converts an incoming spin-$1\over2$ system into an outgoing spin~1
system.

Likewise, Bob's intervention, located space-like with respect to
Alice's, outputs a spin~2 particle when Bob receives one with
spin~$1\over2$. How shall we describe the sequence of events in the
frame where Alice is the first one to act, and in the frame where Bob is
first?

Alice's $A_{\mu m}$ matrices are direct products of a matrix of
dimensions $3\times2$ and the two-dimensional unit matrix, as in
\Eq{dirprod}). Thereafter, there is a free unitary evolution, where
$U_{BA_\mu}$ has rank~6. Then Bob's $B_{\nu n}$ matrices are direct
products of a 3-dimensional unit matrix and one of dimension
$5\times2$.  The final $\rho$ is 15-dimensional (the final quantum
system consists of a particle of spin~1 and a particle of spin~2). A
similar description holds, mutatis mutandis, in the frame where Bob acts
first (this frame is denoted by primes). The unitary matrix
$V'_{AB_\nu}$ for the free evolution from \sfB\ to \sfA\ is of order 10,
while $U_{BA_\mu}$ was of order~6. Obviously these cannot be Lorentz
transforms of each other.  They would not be Lorentz transforms even if
dimensions were the same.  However, the final $\rho_f$ and $\rho'_f$
have to be Lorentz transforms of each other.

Are $A_{\mu m}$ and $A'_{\mu m}$ related by a Lorentz transformation? We
have seen that $A_{\mu m}$ is a direct product of a matrix of dimension
$3\times2$ and the two-dimensional unit matrix. On the other hand,
$A'_{\mu m}$ is a direct product of a matrix of dimension $3\times2$ and
the \mbox{5-dimensional} unit matrix (the latter acts on the spin~2
particle that Bob has produced). Then, the non-trivial parts of $A_{\mu
m}$ and $A'_{\mu m}$, both rectangular $3\times2$ matrices, are Lorentz
transforms of each other. We may also, if we wish, call the complete
$A_{\mu m}$ and $A'_{\mu m}$ matrices ``Lorentz transforms'' if we
accept that unit matrices of any order be considered as Lorentz
transforms of each other.

\bc {\bf IV. SUPERLUMINAL COMMUNICATION?}\ec

Bell's theorem~\cite{Bell} has led some authors to suggest the
feasibility of superluminal communication by means of quantum
measurements performed on correlated systems far away from each other
\cite{Herbert,Garuccio}. It will now be shown that such a possibility is
ruled out by the present relativistic formalism. We have already assumed
that there exists a partial ordering of events. Superluminal
communication would mean that the deliberate choice~\cite{freewill} of
the test performed by an observer (or the random choice of the test
performed by his apparatus) could influence in a deterministic way, at
least statistically, the outputs of tests located at a space-like
distance from that observer (or apparatus) and having a later
time-coordinate. If this were true for any pair of space-like separated
events, this would lead to the possibility of propagating information
backwards in time between events with time-like separation. For example,
we may have \sfA\ in the past light cone of \sfB, and both \sfA\ and
\sfB\ space-like with respect to \sfC. Then \sfB\ could superluminally
influence \sfC\ in the frame where \sfB\ occurs earlier than \sfC, and
in another frame \sfC\ would likewise influence \sfA, so that \sfB\
could indirectly influence \sfA. Therefore the assumption of Lorentz
invariance, and the existence of random inputs, and the restriction of
causal relationships between time-like related events to the future
direction, are incompatible with causal relationships at spatial
distances.

All this was discussed ad nauseam at the classical level many years ago,
when tachyons were popular \cite{tachyon,feinberg}. More recently,
superluminal group velocities have actually been observed in barrier
tunneling in condensed matter \cite{chiao1,chiao2}. However, special
relativity does not forbid the {\it group\/} velocity to exceed~$c$. It
is the {\it front\/} velocity of a wave packet that is the relevant
criterion for signal transmission, and the front velocity never
exceeds~$c$. What novelty does quantum theory bring to this issue? The
common wisdom is that the measuring process creates a ``reality'' that
did not exist objectively before the intervention~\cite{delayed}. Let us
examine this claim more carefully.

Consider a {\it classical\/} situation analogous to the EPRB setup: a
bomb, initially at rest, explodes into two fragments carrying opposite
angular momenta. Alice and Bob, far away from each other, measure
arbitrarily chosen components of ${\bf J}_1$ and ${\bf J}_2$. (They can
measure all the components, since these have objective values.) Yet,
Bob's measurement tells him nothing of what Alice did, nor even whether
she did anything at all. He can only know with certainty what {\it
would\/} be the result found by Alice {\it if\/} she measures her {\bf
J} along the same direction as him, and make statistical inferences for
other possible directions of Alice's measurement.

In the quantum world, consider two spin-$1\over2$ particles in a
singlet state. Alice measures $\sigma_z$ and finds +1, say. This tells
her what the state of Bob's particle is, namely the probabilities that
Bob {\it would\/} obtain +1 {\it if\/} he measures (or has measured, or
will measure) {\boldmath $\sigma$} along any direction he chooses. This
is manifestly counter\-factual information:  nothing changes at Bob's
location until he performs the experiment himself, or receives a
classical message from Alice telling him the result that she found. No
experiment performed by Bob can tell him whether Alice has measured (or
will measure) her half of the singlet.  The rules are exactly the same
as in the classical case. It does not matter at all that quantum
correlations are stronger than classical ones and violate the Bell
inequality.

A seemingly paradoxical way of presenting these results is to ask the
following naive question: suppose that Alice finds that $\sigma_z=1$
while Bob does nothing. When does the state of Bob's particle, far away,
become the one for which $\sigma_z=-1$ with certainty? Though this
question is meaningless, it has a definite answer: Bob's particle state
changes instantaneously. In which Lorentz frame is this instantaneous?
In {\it any\/} frame! Whatever frame is chosen for defining
simultaneity, the experimentally observable result is the same, owing to
\Eq{etcr}). This does not violate relativity because relativity is built
in that equation, as will now be shown in a formal way.

Consider again Eqs.~(\ref{rhof}) and (\ref{K}) which give the final
(unnormalized) $\rho_f$ following two interventions in which Alice gets
the result $\mu$, and then Bob gets the result $\nu$. The probability
for that pair of results is $\Tr(\rho_f)$. If event \sfB\ lies in the
future light cone of \sfA, there can be ordinary classical communication
from \sfA\ to \sfB\ and there is no causality controversy. We are
interested here in the case where \sfB\ is spacelike with respect to
\sfA. The problem is to prove that the probability of Bob's outcome
$\nu$ is independent of whether or not Alice intervenes before him (in
any Lorentz frame). Note that the unitary matrices in \Eq{K}) are the
Green's functions for the propagation of the {\it complete\/} quantum
system, and that its subsystems may interact in a nontrivial way even
when they are macroscopically separated (for example, these may be
charged particles).

Fortunately, we don't need to know these Green's functions explicitly.
We simply note that the probabilities that we are seeking are invariant
under unitary transformations of the various operators in Eq.~(\ref{K}).
In particular, they are not affected by the initial $U_{A0}$ and final
$U_{fB_\nu}$. There still is the intermediate unitary operator
$U_{BA_\mu}$ for the propagation of the composite quantum system between
times $t_A$ and $t_B$. That quantum system is not a localized object.
Its velocity is not a well defined concept and it is meaningless to
argue that it is less than the velocity of light. However, it is
possible to eliminate $U_{BA_\mu}$ by using the same stratagem as in
Sect.~III: we perform a Lorentz transformation of the spacetime
coordinates, which is implemented by a unitary transformation of the
quantum operators (so that all probabilities are invariant), in such a
way that the time elapsing between interventions \sfA\ and \sfB\ is
arbitrarily small, and therefore $U_{BA_\mu}\to{\bf1}$.

The probability that Bob gets a result $\nu$, irrespective of Alice's
result, thus is

\beq p_\nu=\sum_\mu\Tr\Bigl(\,\sum_{m,n}B_{\nu n}\,A_{\mu m}\,\rho\,
  A^\dagger_{\mu m}\,B^\dagger_{\nu n}\Bigr). \label{pnu} \eeq
We now employ \Eq{etcr}) to exchange the positions of $A_{\mu m}$ and
$B_{\nu n}$, and likewise those of $A^\dagger_{\mu m}$ and
$B^\dagger_{\nu n}$, and then we move $A_{\mu m}$ from the first
position to the last one in the product of operators in the traced
parenthesis. We thereby obtain expressions

\beq \sum_m A^\dagger_{\mu m}\,A_{\mu m}=E_\mu. \eeq
As explained in~\cite{I}, these are elements of a positive operator
valued measure (POVM) that satisfy $\sum_\mu E_\mu={\bf1}$. Therefore
\Eq{pnu}) reduces to

\beq p_\nu=\Tr\Bigl(\sum_n B_{\nu n}\,\rho\,B^\dagger_{\nu n}\Bigr),\eeq
whence all expressions involving Alice's operators $A_{\mu m}$ have
totally disappeared. The statistics of Bob's result are not affected at
all by what Alice may do at a spacelike distance, so that no
super\-luminal signaling is possible.

Note that in order to obtain meaningful results the entire experiment
has to be considered as a whole: namely, what was prepared in the past
light cone of {\it all\/} the interventions, and the complete set of
results that were obtained, and are known in their joint future light
cone. It is tempting and it is often possible to dissect an experiment
into consecutive steps, just as it is often possible to discuss
separately the properties of entangled particles.  However, if
ambiguities (or conflicting predictions, or any other ``paradoxes'') are
encountered, what has to be done is to consider the whole entangled
system and the whole experiment.  Contrary to naive intuition, there is
no physical state vector that interpolates between the initial and the
final states.  Such interpolations can formally be written, but they are
not unique, not Lorentz covariant, and therefore they are physically
meaningless.

Yet, there is an important exception to the above rule: if there exists
a spacetime point such that there are interventions in the past and
future light cones of that point, but no intervention is spacelike with
respect to it, then it is possible to divide the experiment into two
steps, before and after that point. It is then meaningful to define not
only an initial state $\rho_0$ and a final state $\rho_f$, but also an
intermediate state $\rho_i$ {\it at that point\/}. It is conventional to
refer such a state to a spacelike hyperplane that passes through the
point, but actually the only role of that hyperplane is to define the
Lorentz frame in which we write a mathematical description of the
state.

It thus appears that the notion of quantum state should be reassessed.
There are two types of states: first, there are physically meaningful
states, attached to spacetime points with respect to which no classical
intervention has a spacelike location. Then, between any two such
points, we may draw a continuous timelike curve and try to attach a
quantum state to each one of the points of that curve. These
interpolating states can indeed be defined as shown in the present
article, by considering a set of parallel spacelike hyperplanes.
However, states defined in such a way are merely formal mathematical
expressions and they have no invariant physical meaning.

In summary, relativistic causality cannot be violated by quantum
measurements. The fundamental physical assumption that was needed in the
above proof was that Lorentz transformations of the spacetime
coordinates are implemented in quantum theory by {\it unitary\/}
transformations of the various operators. This is the same as saying
that the Lorentz group is a valid symmetry of the physical system.

\bc {\bf V. CONCLUDING REMARKS}\ec

In the present article it has been shown that a careful treatment,
avoiding any speculations that have no experimental support, leads to
the ``peaceful coexistence''~\cite{shimony} of quantum mechanics and
special relativity. The spacetime coordinates of the observers'
interventions are classical parameters subject to ordinary (classical)
Lorentz transformations. The latter are implemented in quantum mechanics
by unitary transformations of the operators. There are no essentially
new features in the causality issue that arise because of quantum
mechanics. Quantum correlations do not carry any information, even if
they are stronger than Bell's inequality allows. The information has to
be carried by material objects, quantized or not.

The issue of information transfer is essentially nonrelativistic.
Replace ``superluminal'' by ``supersonic'' and the argument is exactly
the same. The maximal speed of communication is determined by the
dynamical laws that govern the physical infrastructure. In quantum field
theory, the field excitations are called ``particles'' and their speed
over macroscopic distances cannot exceed the speed of light. In
condensed matter physics, linear excitations are called phonons and the
maximal speed is that of sound.

The classical-quantum analogy (with bomb fragments carrying opposite
angular momenta ${\bf J}_1=-{\bf J}_2$) becomes complete if we use
statistical mechanics for treating the classical case. The distribution
of bomb fragments is given by a Liouville function in phase space. When
Alice measures ${\bf J}_1$, the Liouville function for ${\bf J}_2$ is
instantly altered, however far Bob is from Alice. No one would find this
surprising, since it is universally agreed that a Liouville function is
only a mathematical tool representing our statistical knowledge.
Likewise, the wave function $\psi$, or the corresponding Wigner
function~\cite{wigner} which is the quantum analogue of a Liouville
function, should be considered as mere mathematical tools for computing
probabilities. It is only when they are regarded as physical objects
that superluminal paradoxes arise.

The essential difference between the classical and quantum functions
which change instantaneously as the result of measurements is that the
classical Liouville function is attached to objective properties that
are only imperfectly known. On the other hand, in the quantum case, the
probabilities are attached to {\it potential\/} outcomes of mutually
incompatible experiments, and these outcomes do not exist ``out there''
without the actual interventions. Unperformed experiments have no
results~\cite{unp}.

\bc {\bf ACKNOWLEDGMENTS}\ec

I am grateful to California Institute of Technology, where this research
program began, for its hospitality, and in particular to Chris Fuchs for
many helpful comments and an inexhaustible supply of references. I also
had fruitful discussions with Dagmar Bru\ss, Rainer Plaga, Barbara
Terhal and Daniel Terno. This work was supported by the Gerard Swope
Fund and the Fund for Encouragement of Research.

\newpage\vspace*{\fill} 

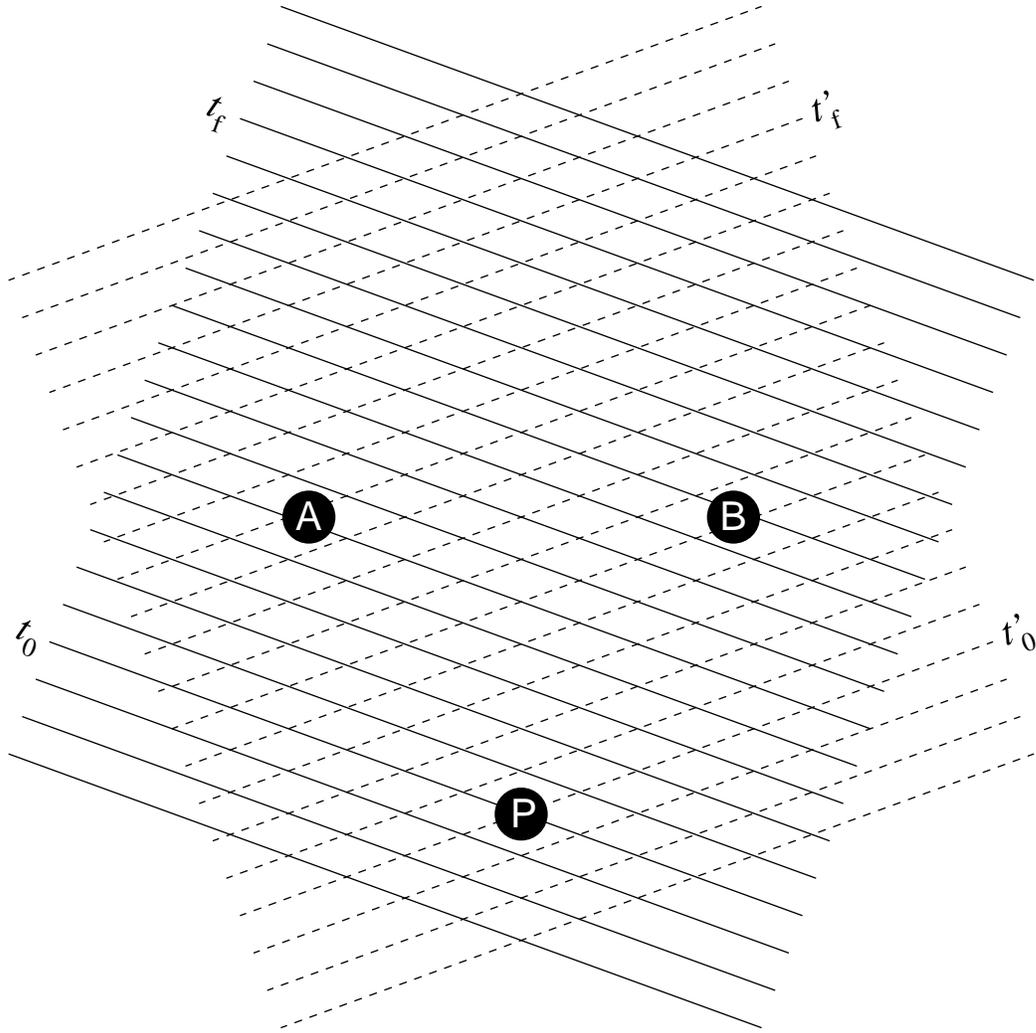
\noindent FIG. 1. \ A quantum system is prepared at point {\sf P}. The
interventions \sfA\ and \sfB\ are mutually space-like. The solid and
dotted lines represent equal times, $t$ and $t'$ respectively, in two
Lorentz frames in relative motion. Event \sfA\ occurs first in $t$-time,
and event \sfB\ is the first one in $t'$-time.


\end{document}